\documentclass[12pt]{article}

\usepackage{graphics}
\usepackage{amsmath}
\def\ni{\noindent}
\def\ph{{\phantom{...}}}
\def\={\phantom{..} = \phantom{..}}

\def\+{\phantom{..} + \phantom{..}}
\def\>{\phantom{..} > \phantom{..}}
\def\<{\phantom{..} < \phantom{..}}
\def\-{\phantom{..} - \phantom{..}}

\def\all{\phantom{..} \hbox{for all} \phantom{..}}

\def\bq{\begin{quote}}
\def\eq{\end{quote}}
\def\be{\begin{equation}}
\def\ee{\end{equation}}
\def\bar{\begin{eqnarray}}
\def\ear{\end{eqnarray}}
\def\no{\nonumber}

\def\iif{\hbox{if}} 
\def\otw{\hbox{otherwise}}

\def\Sch{Schr{\"o}dinger}
\def\Schism{Schr{\"o}dingerism}
\def\Schist{Schr{\"o}dingerist}
\def\Schists{Schr{\"o}dingerists}

\def\Copism{Copenhagenism}
\def\Copist{Copenhagenist}
\def\Copists{Copenhagenists}

\def\Poin{Poincar{\'e}}

\def\sdoic{sensitive dependence on initial conditions}
\def\TDotS{That Dot on the Screen}

\def\cHnl{{\bf{H}_{NL}}}
\def\cH{{\bf{H}}}

\def\siN{\sum_{i=1}^N}

\def\sinn{\sum_{i=1}^n}
\def\sjn{\sum_{j=1}^n}

\def\uy{\underline{y}}
\def\Born{\hbox{Born}}

\title{\bf That Dot on the Screen:\\[0.5in]
 also, what about Born?\\[0.5in] And other objections to wavefunction physics.\\[3in]}

\author{W. David Wick\footnote{email: wdavid.wick@gmail.com}}

\begin{document}
\maketitle
\pagebreak

\section*{Abstract}
In this paper I address the most common objections to the claim that \Sch\ was right in 1926:
the wavefunction provides the correct, and {\em complete}, description of atomic phenomena.
I suggest that 
the line of droplets in the Wilson cloud chamber, the click of the ``photon detector",
and ``that dot on the screen" can all be explained within a context of wavefunction models and
\Sch's-type equations, albeit nonlinear. 
No auxiliary hypotheses about point particles or probabilities are required. 
The random locations of
the triggered ``particle detectors" can be explained by ``chaos" (meaning \sdoic).
Even Born's {\em ad hoc} invocation of probabilities may be justifiable in certain circumstances.
As an illustration, I present simulations from a (toy) wavefunction ``particle-detectors" model.

\section{Introduction}

What about \TDotS?

I suppose it is the first objection most 
people will make, if told that \Sch\ was right about his wavefunction. 
If a single grain on a photographic emulsion can be exposed by a beam of light,   
how can light be fully described by a wave spread over the plate's whole area? 
OK, they admit, the wave is there, but ``the photon" must be too,
perhaps guided by the wave to its destination, like a bottle tossed into the ocean?
Or perhaps the wavefunction is sucked into the grain, at the moment of detection?
 Or is it that
our minds cannot grasp the ineffable conception
of ``wave-particle duality"?

For others, it might be the click of the ``photon detector". But for me, it was the line
of droplets formed in a jug containing some supersaturated vapor, and
equipped with a bit of radioactive
material fixed at the bottom.\footnote{Someone in a lab showed me 
their jury-rigged Wilson cloud chamber,
but I cannot recall where---it might have been at the University of Chicago, 
where I was an undergraduate physics major, or at the University of Washington in Seattle, 
where I was (briefly)
a grad student in the Physics Department, or at Princeton where I was a post-doc for three years.
I apologize for not remembering who constructed it or where; but I offer a delayed thanks.}  

I should first clear up certain confusions about \Sch's theory and about ``particle detectors".
In 1929 a \Schist\ in England, Nevill F. Mott, discussed, \cite{Mott}, 
why the new wavefunction theory
was not instantly refuted by those lines appearing in cloud chambers. 
Do not imagine a wave representing the alpha-particle, wrote Mott, trapped originally in the 
nucleus but then ``leaking out". \Sch's wavefunction is not a wave 
in three-dimensional physical space, he points out, 
but in an abstract space of 3N dimensions, which results from consideration
of everything in the chamber---the nucleus, the alpha particle, 
the water (or other vapor) molecules, etc.
These vapor molecules serve as the nucleating centers of droplets, representing the
detectors in this scenario. 

Mott also addressed the question: how can such a wavefunction generate a line, 
as opposed to a spherical wave, of water droplets? 
He argued\footnote{On the basis of 2nd-order perturbation theory,
an approach which can be criticised as not definitive.} that if two vapor molecules become ionized,
and can therefore instigate droplet formation, they are likely to lie on a straight line
passing through the nucleus. However, Mott did not explain 
why we often see a single line---as opposed to a spray of lines.

Other so-called ``particle detectors" include grains on a photographic film, electrical devices
(``photon detectors"), and even the rod-and-cone molecules in your retina. In each case,
the detector works like this: some energy is deposited in the detector, which triggers 
an amplification step, leading to a macroscopic change in something we can perceive.
For example, on the photographic film light energy absorbed by a grain leads to a deposit of
silver, which, on development of the film, leads to a darkened spot. In the eye,
absorption of light energy leads to a conformational change in a molecule (retinal)
which then pushes on a larger molecule (rhodopsin), resulting in an electrical signal stimulating
a neuron, etc., etc., until the lips and the vocal cords announce: ``I see something!".
  
In an experiment, a beam is directed at the apparatus. Some time is alloted for detection
(after which the detector may have to be reset to its ``untriggered" state). Depending on the
incoming flux and the window of detection, the result may be one detector triggered, 
or many, or none, in that ``run" of the experiment.
Of course, what excites our interest is that first case: one detection. One dot on the screen. 
One click of the ``photon detector". One line of water droplets in that jug. 
That is what we must explain in wavefunction theory.

\Sch's wavefunction is a complex-valued function, having ``real" and ``imaginary" parts,
although we can drop this 19-th Century 
phraseology (with its mystical connotations) and simply remark
that it has two (real-valued) components. Right at the beginning, there arose the problem
of giving the wavefunction a physical interpretation. Then, in late 1926, Max Born 
(one of Heisenberg's collaborators) had an epiphany: 
in an article about scattering by a nucleus, \cite{Born}, he suddenly
asserted that ``only one interpretation [of the wavefunction] is possible": 
meaning as a probability.
In a footnote, he wrote: ``More careful consideration shows [it must be] proportional to
the square [of
the wavefunction]." 
Of course, neither is sensible, as a complex (or two-valued) number 
cannot be recast as a probability, nor can its square, which might be negative. 
The \Copists\ soon repaired it to
the claim that the modulus squared ($|\psi|^2 = \psi^*\psi$) is a probability density. 
At least it is non-negative and its integral over all space is conserved under the time-evolution,
 and can be set equal to one if desired. 
But they gave no rational justification
for this interpretation. 

Probability entered ``quantum" physics in this {\em ad hoc} way.
At first, supporters said that the integral of $\psi^*\psi$ over the top of a box
gives ``The probability the particle {\em is} in the top part"; 
but after Einstein, Podolsky and Rosen in 1935, \cite{EPR},  fired their broadside 
against the claim that the theory gives a complete description of 
particle positions and momenta, they dropped the ``is" in favor of ``finding": the integral
gives the probability of {\em finding} the
particle in the top half of the box (if, I suppose, it were
to be observed).\footnote{John Bell remarked upon the \Copists\ unwillingness to say
what is {\em beable}, as opposed to what is {\em findable}, \cite{Bell}. Of course,
we \Schists\ regard the wavefunction as the {\em beable}.}

In addition to the interpretation issue,
\Sch's theory presented two additional conundrums, for neither of which did he find a resolution: 
``cats" and randomness. \Sch\ himself identified the first difficulty in his 1935
``cat" paper, \cite{catpaper}. Unfortunately, he embodied the problem in a Rube-Goldberg-complicated
metaphor about a cat locked in a steel box equipped with a Geiger counter, a bit of uranium,
a hammer, and a flask of prussic acid. Perhaps a uranium nucleus decayed, the counter clicked,
the hammer swung shattering the flask, and the cat died. Or perhaps not. This presentation
confounded the issue he wanted to illustrate---that in the wavefunction theory the cat gets 
smeared out between two alternatives (life or death)---with the subjective theory of knowledge
or probability. (``Before I open the box, I don't know whether the cat is alive or dead!" So?)
The cleaned-up modern version loses the steel box and the rest of the apparatus
except for the Geiger counter and the uranium. 
Now the cat either is startled by the counter's click and jumps, say, one meter to the left, 
or there was no click and the cat sits still. 

The problem for wavefunction theory is that
you don't get the ``or" in the last sentence; you get an ``and". 
More formally: $\psi$ is a superposition 
of a wave packet representing ``counter clicked, cat leapt"
{\em and} a wavepacket representing ``no click, no jump". As von Neumann demonstrated in
his book of 1932, \cite{vonNeumann}, nothing exists in the theory to eliminate 
one of these wave packets and leave the other.  

In today's physics,
by a ``cat" we mean a macroscopic object whose dispersion of its center of mass,
as computed from the wavefunction, is larger than the object's size. Measurement situations
(we have one in the leaping-and-not-leaping cat, because the cat's position ``measures"
the state of the uranium nucleus) always produce some kind of cat, in linear wavefunction
theory. This anomaly became known as the Measurement Problem.

The other conundrum concerns the fact that, in certain experiments, no matter how much
effort is expended on reproducing initial conditions between ``runs", the outcome varies
randomly. But \Sch's theory was completely deterministic: if you can perfectly control
the initial wavefunction, whatever you observe at a later time is predetermined.

In a series of papers beginning in 2017, this author proposed solutions for these difficulties.
For the MP he suggested modifying \Sch's linear equation of 1926. Desiring to stay
within the physics tradition, he  
preserved the Hamiltonian structure but added a 
term which became known as ``WaveFunction Energy" (WFE). 
The novel term was quartic rather than quadratic in $\psi$, 
implying that the induced dynamics is nonlinear. Superposition is broken
at the macroscopic, but not the microscopic, level. As in \Sch's original theory, 
the time evolution
preserved the total energy exactly, as well as the norm of the wavefunction.
Cats were blocked by an energy barrier. All this is explained in paper I, \cite{WickI}.

In a second paper, \cite{WickII}, he took on the determinism problem 
by suggesting that there should
be a universal, random component of wavefunctions, and showed that such a theory can
generate a violation of Bell's Inequality. However, noting that high-dimensional,
nonlinear evolution equations often display the phenomena dubbed ``chaos", he dropped
the idea and in a third paper, \cite{WickIII},
 provided some evidence ({\em via} simulating a toy model) for 
\sdoic\ as an explanation of random outcomes in physics experiments. This paper
also contained a possible criterion for the onset of chaos, involving a determinant
(thus the ``Determinant Criterion for Instability", or DCI).
In a later paper, \cite{Wickchaos},
 he made a closer analysis of the DCI and provided some evidence, 
again from simulating
a small (``three-qubit") model, that the DCI was indeed the threshold of chaos.

Thus, in this proposal, nonlinearity explained the absence of cats, why classical
physics works so well at the macro level, and the inevitable appearance of randomness in
``quantum" physics experiments.\footnote{Concerning the quoted word, which did appear in
his first four papers on the subject, this author has since dropped it from his vocabulary,
as nothing discrete or ``quantized" is involved in his proposed explanations.}
This field joins the others of science, in which randomness is explained along the lines
proposed by Henri Poincar{\' e} in 1896, \cite{Poincare}: 
due to instabilities in the dynamics, and our inability
to precisely control all initial conditions. 

Returning to that dot on the screen: in what scenarios do we expect it to appear? The simplest
hypothesis is that the energy available in the system is adequate only to trigger one
detector (develop one grain,...). That will prevent a sea of dots. 
Assuming this bound on the energy, we are still faced with the paradox of cats,
here meaning detectors that are both triggered and untriggered. The nonlinear terms
in the modified \Sch's equation should take care of that. 

One solved (or at least simulated) model being worth a thousand words, I produce one,
described in section \ref{modelsection}. Since, unfortunately, I did have recourse to the
computer and simulations, and since simulating wavefunction models in the continuum
is virtually impossible, I was limited to models with a discrete ``spatial' component,
 and to an extremely 
simplified version of a detector. (Computer run time is the limiting factor; discussed in section
\ref{QandAsection}.) In a separate section, I discuss a question about WFE neglected
in previous work: should distinguishable subsystems have their own WFEs? The answer will
inform the choice of WFE in the model.   

The model, even ignoring simulation results,
will serve to render some points in the discussion more concrete. 
The thousand words come afterwards. 
As I suspect the 
reader will have formulated by then many objections, 
I will adopt  
a ``Question-and-Answer" (Q\&A) format. Please restrain your outrage
until you arrive at that section.

In the final section, I make some remarks about \Poin, Born, and why the word
``particle" sometimes appeared in previous sections. Technical matters are explained in 
a Statistical and Computational Appendix.

\section{The Detector Model\label{modelsection}}

The detector model will be defined by variables `$x$', the ``particle position",
taking on possible values: $1,2,..,n$; and `$y_i$', for $i = 1,2,..,n$, the ``detector variables",
taking on
possible values $0,1,...,d$. (I will usually take $d=1$, yielding the interpretation:
$y_i = 0$ means the $i$-th detector was not triggered, and $y_i=1$ means it was triggered.)
The wavefunction is therefore a function of $n+1$
arguments:

\be
\psi(x,y_1,y_2,...,y_n) \= \psi(x,\uy),
\ee

\ni where I have introduced the symbol $\uy$ for the collection of detector variables.

The Hamiltonian\footnote{For \Copists, the ``Hamiltonian" is taken to be
a linear Hilbert Space operator, and what we \Schists\ mean by the term 
is called ``the expected energy",
because the former group believes that energy of an atom takes discrete values and
the system always lies (will be found to lie?) 
in an eigenstate of that operator. \Schists\ of course don't accept \Copist\
ideology; for us, the Hamiltonian is a function of the state, namely $\psi$, as in
classical mechanics, where the state is a set of P's and Q's.}  will be:

\bar
\no \cH &\=& <\psi\,|\,K\,|\,\psi> \+ \alpha\,<\psi\,|\,\sinn\,\delta_i(x)\,(\,y_i - d\,)^2\,
|\,\psi> \+\\
\no && v\,<\psi\,|\,\sinn\,y_i\,|\,\psi> \+ \cHnl.\\
&&\label{Ham}
\ear

\ni Writing this out, putting in the variables:

\bar
\no \cH &\=& \sum_{i,\uy}\,\sum_{i',\uy'}\,\psi^*(i,\uy)\,K_{i,\uy;i',\uy'}\,\psi(i',\uy')
 \+ \alpha\,\sum_{i,\uy}\,|\psi|^2(i,\uy)\,(\,y_i - d\,)^2 \+\\
\no && v\,\sum_{i,\uy}\,|\psi|^2(i,\uy)\,\sjn\,y_j \+ \cHnl. \\
&&\label{Hamtwo}
\ear

\ni Here, $\alpha$ and $v$ are positive constants associated to ``particle-detector
interaction" and ``detector energy", respectively.
The first term is supposed to be a mock ``kinetic energy of the particle" plus a ``kinetic energy
of a detector pointer";
the matrix $K$ is chosen so that the first term in (\ref{Hamtwo}) equals:

\def\uz{\underline{z}}

\bar
\no && \left(\,\frac{1}{2\,\hbox{M}}\,\right)\,\sum_{\uy}\,\sum_{i,j:|i-j|=1}\,
|\,\psi(i,\uy) -\psi(j,\uy)\,|^2 \+\\
\no && \left(\,\frac{1}{2\,\hbox{DM}}\,\right)\,\sum_{i=1}^n\,\sum_{\uy,\uz: |\uy - \uz|=1}\,
|\,\psi(i,\uy) -\psi(i,\uz)\,|^2.\\
\no && |\,\uz - \uy \,| \= \sum_{i=1}^n\,|\,y_i - z_i\,|.\\
&&
\ear

\ni `M' is supposed to be the `particle' mass and DM the `detector' mass (think of a part of
the detector that moves, perhaps like a needle on a scale).

The last term in (\ref{Ham}) is the ``nonlinear energy" and involves WFE,
to be described in section \ref{subsystemssection}.

The evolution equation has the form:

\be
\sqrt{-1}\,\frac{\partial \psi}{\partial t} \= \frac{\partial \cH}{\partial \psi^*};
\ee

\ni equivalently, introduce $P(i,\uy)$ and $Q(i,\uy)$ equal to the real and imaginary
parts of $\psi(i,\uy)$ respectively and write Hamilton's equations. 
Without the nonlinear term, the equation is easily seen 
to be \Sch's (in units where $\hbar = 1$).

The initial condition was chosen to be:

\bar
\no P_{i,\uy} &\=& \begin{cases}
 \sqrt{1 - \epsilon^2}\,\delta_{0,0} \+ \epsilon\,\cos(\theta)\,z,& \iif\ph y_j < d, 
 \all j = 1,..,n;\\
 0,& \otw.
 \end{cases} \\
 \no Q_{i,\uy} &\=& \begin{cases}
 \epsilon\,\sin(\theta)\,z,& \iif\ph y_j < d, 
 \all j = 1,..,n;\\
 0,& \otw.
 \end{cases}\\
&&
\ear

\be
\psi_{i,\uy} = P_{i,\uy} \+ \sqrt{-1}\,Q_{i,\uy}.
\ee

Here the $z$ and $\theta$ were choosen randomly and independently for each $(i,\uy)$
with $z \sim N(0,1)$ and $\theta \sim U(0,2\pi)$; then the wavefunction was normalized
to make its absolute values sum to one.

The meaning of this is: for $\epsilon = 0$ the initial wavefunction is concentrated
entirely on the configuration: $(0,0,...,0)$, i.e., with the ``particle at the origin"
and all detectors untriggered; with $\epsilon >0$ the wavefunction 
is nonvanishing for some spread in ``particle location" but still no detectors are triggered.

The observables\footnote{Please defer your objections here to the Q\&A, where I will
discuss philosophy of measurement and of modeling.}  will be the $n$-tuple:
\def\Ob{\hbox{Ob}}

\bar
\no \Ob_i &\=& <\psi\,|\,y_i\,|\,\psi> \\
\no &\=& \sum_{j,\uy}\,|\psi(j,\uy)|^2\,y_i.\\
\no i &\=& 1,2,...,n.\\
&&
\ear

Also of interest is the $n$-tuple of probabilities I will call ``Born":

\bar
\no \Born_i &\=& <\psi\,|\,\delta_i(x)\,|\,\psi> \\
\no &\=& \sum_{\uy}\,|\psi(i,\uy)|^2.\\
\no i &\=& 1,2,...,n.\\
&&
\ear

\section{Distinguishable Subsystems and WFE\label{subsystemssection}}

Wavefunction Energy in previous work took the form:

\be
w\,\left\{\,<\psi\,|\,S^2\,|\,\psi> - \left[\,<\psi\,|\,S\,|\,\psi>\,\right]^2\,\right\},
\label{expression}
\ee

\ni where `$w$' is a positive constant,$\psi$ denotes the wavefunction, and `$S$'
stands for an operator created out of spatial or momentum variables.  In a one-spatial-dimensional
model with spatial WFE we may take:

\be
S \= \siN\,x_i.
\ee

This WFE can also be written:

\be
w\,N^2\,D(\psi),
\ee

\ni where $D(\psi)$ is the (squared) dispersion of the Center-of-Mass (CoM):

\bar
\no D(\psi) &\=& <\psi\,|\,\hat{S}^2\,|\,\psi> - \left[\,<\psi\,|\,\hat{S}\,|\,\psi>\,\right]^2;\\
\no \hat{S} &\=& \left(\,\frac{1}{N}\,\right)\,\sinn\,x_i.\\
&&
\ear

Now let us consider a system formed from the union of two subsystems, say with arguments
$x_1^{(1)},x_2^{(1)},...,x_N^{(1)}$ and  
$x_1^{(2)},x_2^{(2)},...,x_N^{(2)}$. Should we lump the subsystems together, ignore the
superscripts, and use expression (\ref{expression}) with $N$ replaced by $2\,N$? Or
should we write

\be
\hbox{WFE} \= \hbox{WFE}_1 \+ \hbox{WFE}_2,
\ee

\ni with

\bar
\no \hbox{WFE}_j &\=& 
w\,\left\{\,<\psi\,|\,S_j^2\,|\,\psi> - \left[\,<\psi\,|\,S_j\,|\,\psi>\,\right]^2\,\right\};\\
\no S_j &\=& \siN\,x_i^{(j)};\\
&&
\ear

\ni for $j = 1,2$? I presume the deciding consideration is distinguishability of the two
subsystems.

Indistinguishability of particles, say of electrons in an atom, has of course played
a celebrated role in the history of atomic physics and chemistry. \Sch's equation plus
Pauli's Exclusion Principle for indistinguishable electrons in an atom suffice as the
explanation of the Periodic Table, the chemical bond,
 and much else in chemistry. But what about electrons
in atoms which are part of two separated devices? Surely they are distinguishable as resident
in different locations? 

The author confronted this question in paper I, imagining the rather implausible set-up
of an EPRB experiment in which Alice's and Bob's detectors have needles that can move
up or down. If the devices are identical and perfectly oriented to measure along the same
direction, they are presumed to obtain perfectly anti-correlated outcomes; i.e., ``needle went up"
in Alice's lab always means ``needle went down" in Bob's. If we had assumed WFE on the joint CoM
of the two needles, it might not prevent cats, as the CoM might be initially, and remain,
zero (with suitable origin of coordinates). 

I dismissed this objection as requiring a ``knife-edge" coincidence which could never be
arranged. I now think this resolution overlooks an important issue, that of distinguishability.
Hence I propose that the second option for WFE would apply in this thought experiment,
and prevent cats from forming in both laboratories.

For the model, I presumed each detector, located at possible $x$-position `$j$', and having
registration variable $y_j$, is separated from the other detectors and is moreover a
macroscopic device to which WFE should be applied. Following the above argument, I
propose that WFE should be a sum of separate WFE$_j$'s which involve the ``marginal" 
distributions only. Hence:

\be
\hbox{WFE} \= w\,\sjn\,\left\{\,\sinn\,\sum_{\uy}\,|\psi|^2(i,\uy)\,y_j^2 - 
\left[\,\sinn\,\sum_{\uy}\,|\psi|^2(i,\uy)\,y_j\,\right]^2\,\right\}.
\ee

\section{Dots and Cats\label{DaCsection}}

With the model now fully specified, we can see more concretely when dots or cats may appear.
Let's set $d = 1$, so that the detectors can take only two values:
$y_j = 0$ means $j$-th detector untriggered, and $y_j=1$ means it was triggered.
In terms of what is observed, namely the vector $\Ob$, 
the detector at position `$i$' was presumed triggered if
$\Ob_i > \delta$. 
Finally, let ``DE", short for Detector Energy, stand for the third term in 
(\ref{Ham}) and (\ref{Hamtwo}). 
Consider three conceivable final-time wavefunctions:

(a) For all $i = 1,2,,,n$:

\be
\psi(i,\uy) \= 
\begin{cases} 
\frac{1}{\sqrt{n}},& \iif\ph \uy = (1,1,...,1);\\
0,& \otw. 
\end{cases}
\ee

\ni Here all detectors are triggered; $DE \= v\,n$; and $\Ob = (1,1,1)$. Not a cat.
(b) For some $k$, $1 \leq k \leq n$:

\be
\psi(i,\uy) \= 
\begin{cases} 
1,& \iif\ph i = k,\ph y_k =1,\ph \hbox{and}\ph y_j=0\ph \iif \ph j \neq k ;\\
0,& \otw. 
\end{cases}
\ee

\ni Here one detector is triggered; $\Ob = (0,0,...,1,...,0)$ with the one in slot `$k$';
$DE = v$. Not a cat.

(c) 

\be
\psi(i,\uy) \= 
\begin{cases} 
\frac{1}{\sqrt{n}},& \iif\ph y_i = 1 \ph\hbox{and}\ph y_j = 0 \ph \hbox{for all}\ph j \neq i;\\
0,& \otw. 
\end{cases}
\ee

\ni Here $\Ob = (1/n,1/n,...)$ and so either all or no detectors are observed
to be triggered, depending on how `$\delta$' was chosen; $DE = v$. The state is a cat.

An initial bound on the energy could eliminate case (a), but not case (c). The latter
case can be prevented by WFE; indeed,

\bar
\no \hbox{WFE} &\=& w\,\sjn \left\{\,\frac{1}{n} - \frac{1}{n^2}\,\right\}\\
\no &\=& w\,\left(\,1 - \frac{1}{n}\,\right).\\
&&
\ear

Thus, a bound on the energy plus a sufficiently large value for parameter `$w$'
will disabuse the cat, leaving only (b) as a possible endpoint.

Of course, it might be that, with such parameter choices, nothing happens.
This is where simulations are needed.
\pagebreak

\section{Results from Simulations\label{simsection}}

Table \ref{table1} shows the parameters used for the simulations.
(FT = Final time; TS = time-steps.) The parameters were chosen by experimentation,
with some goals in mind. First, there should be some wavefunction spread over spatial positions
(``movement of the particle") and detections (``some detectors should be triggered").
As for WFE, controlled by parameter `w', the computed energy
with the parameters in the Table and the initial conditions 
described in section \ref{modelsection}, came out to around 20; hence the choice made was to
render the cat state (section \ref{DaCsection}) almost impossible.

Table \ref{table2} shows the results of a small simulation study.
What is the explanation for the peculiar 
finding that on the basis of one statistic (called `S', see the Statistical and Computational
Appendix) we {\em cannot} reject the Born hypothesis, while on the basis of another
(called `R') we {\em must}? The first conclusion---failure to reject---reflects that the 
outcomes did not have too many improbable cases, i.e., cases in which a dot was
detected at a position while the Born probability of that position was small. 
The second---rejection---reflects an improbable result with too many cases 
in which the location of the detection
was the same as that position where Born was maximal (i.e., too little variance\footnote{As
Fisher famously detected in Mendel's pea-plants data, using his $\chi^2$ test. 
He concluded that the gardiner had ``improved" the outcomes.}). 

Manually inspecting the outcomes---the Born and Ob vectors generated in each run---reveals
the explanation: when there was only one dot (one detector triggered), as learned from Ob,
the triggered detector sat right on the spot of maximal Born probability. 
Which is not in concordance
with Born's hypothesis. See Figure 1.

What was the role of WFE, controlled by parameter `$w$', in these outcomes?
I reran the study using parameters from Table 1, but setting $w=0$.
The result was that, out of 200 runs, no single Dot-on-the-Screen appeared (in fact, no detectors
triggered at all).
Examination of vectors Ob and Born showed that they were almost flat.

Also interesting, of course, is whether, due to the nonlinearity introduced by WFE,
the system is ``chaotic". While I did not make a formal study of this question,
as I did for a different model in \cite{Wickchaos}, I looked for regions of instability
(where det M $< 0$). There were none among the initial configurations as used.
But in that earlier paper I noted that the system entered and departed regions of
instability. This phenomenon appeared also for this model; see Figure 2. ($\hbox{slog}(x) =
\hbox{sign}(x) \log(|x|)$.)

Figure 3 shows two histograms of dots at two final times. I do not know why the system
discriminated against the dot appearing at position 2.0.
 
\begin{table}
\centering
\caption{Model Parameters}
\label{table1}
\begin{tabular}{c c c c c c c c c c}
\hline \hline
n & d & w & v & $\alpha$ & M & DM & $\delta$ & FT & TS \\
\hline
3 & 1 & 20.0 & 10.0 & 15.0 & 0.2 & 0.5 & 0.7 & 20.0 & 20,000 \\
\hline
\end{tabular}
\end{table}

\begin{table}
\centering
\caption{Simulation Results, using parameters from Table \ref{table1}}
\label{table2}
\begin{tabular}{c c c c c}
\hline \hline
Runs & one dot & no dots & p-value (S) & p-value (R)\\
\hline
200 & 77--85 & 115-122 & $> .99999$ & $< .00000$ \\
\hline
\end{tabular}
\end{table}

\begin{figure}
\rotatebox{0}{\resizebox{5in}{5in}{\includegraphics{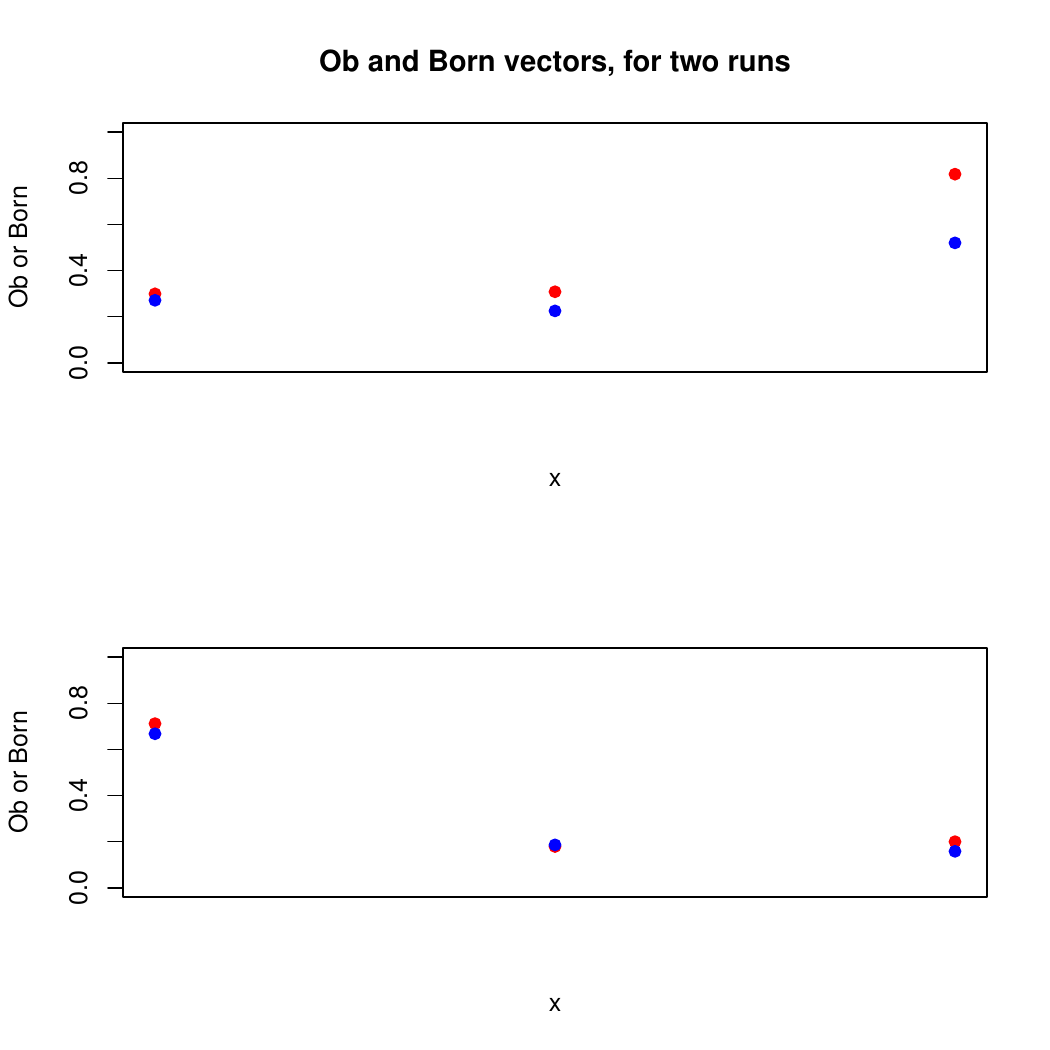}}}
\caption{Results for two typical runs, using parameters from Table 1. 
          Ob in red; Born in blue.}\label{Fig1}
\end{figure}

\begin{figure}
\rotatebox{0}{\resizebox{5in}{5in}{\includegraphics{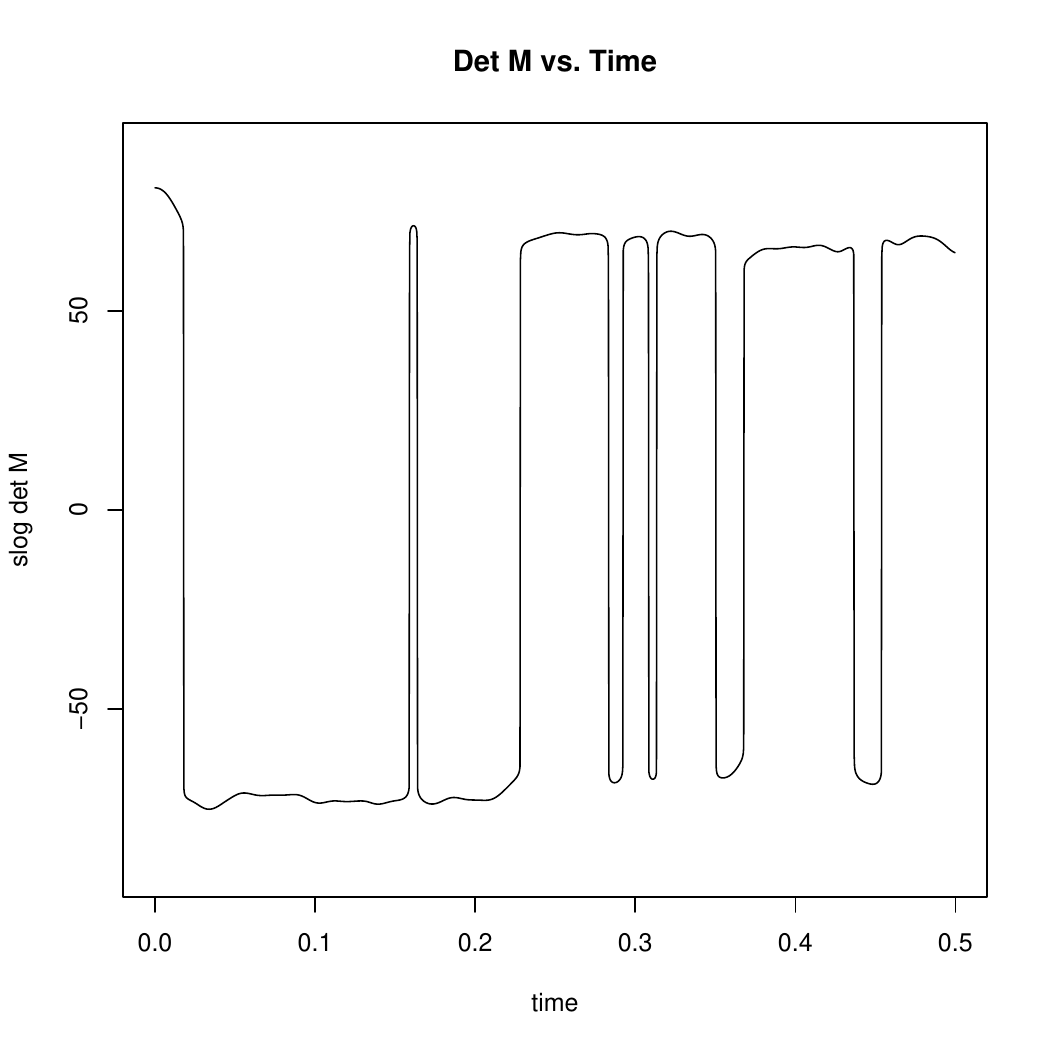}}}
\caption{Plot of det M {\em vs.} time, with parameters from Table 1. 
          }\label{Fig2}
\end{figure}

\begin{figure}
\rotatebox{0}{\resizebox{5in}{5in}{\includegraphics{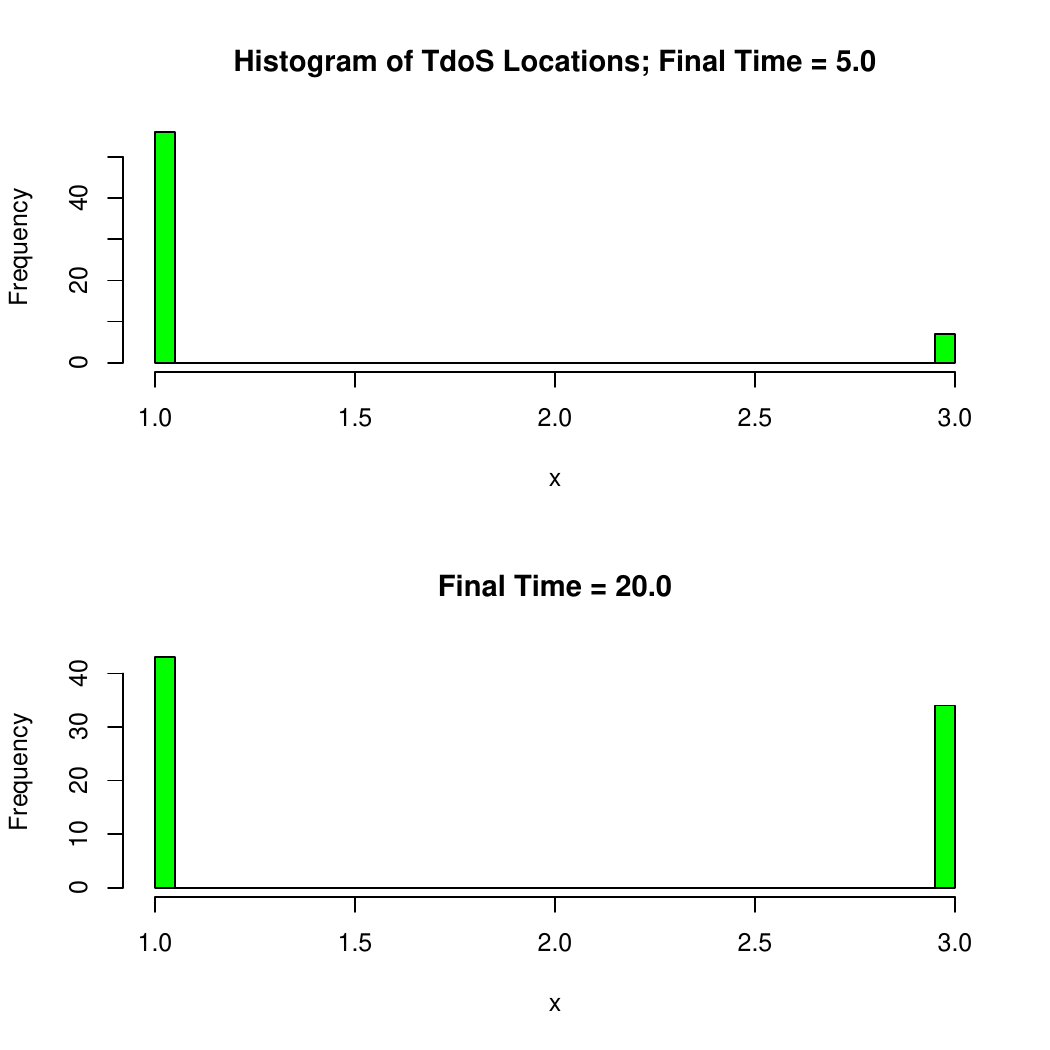}}}
\caption{Histograms of That Dot on the Screen 
          }\label{Fig3}
\end{figure}

\pagebreak
\pagebreak

\section{Q\&A\label{QandAsection}}
\def\bQ{\ni {\bf Question}: }
\def\bA{\ni {\bf Answer}: }

\bQ Your ``observables" appear to take on a continuous range of values. But did not
von Neumann tell us that measurement outcomes lie on a calculable, discrete list of real numbers?

\bA He did. But that claim would astonish anyone who works in a laboratory, or has
dealt with real data. Of course, \Copists\ have their excuses ready: 
if, say, in measuring the energy of an atom, 
the number found doesn't agree with any of the eigenvalues of the Hamiltonian operator,
that is because of the ``Time-Energy Uncertainty Principle". But no such ``Principle"
exists in the theory from the '20s, at least not as a facile reinterpretation of a lemma
in Fourier Analysis, as for the ``Position-Momentum Uncertainty Principle" of Heisenberg.
Rather, believers in the former ``Principle" must derive it from a bespoke analysis
of the measurement context, amounting to a repudiation of the original claim.
  
In my philosophy of measurement, a variety of ``instrumentalism", it suffices to
state which components of your theory (of system+apparatus) constitute the observables.
Then your theory makes a definite prediction, subject to falsification by experiment.
(And I do not accept ``uncertainty" as a legitimate physics word. It is a borrowing
from psychology or insurance or sports betting, and should have no presence in physics.)

\bQ I'm not impressed by your ``toy" model with three possible positions for the particle
and two states for each detector. Even my three-year-old would not be interested.

\bA Give me a break. The model with `$n$' positions for the particle and `$d$' for the detectors
has dimension (meaning components of the wavefunction): $2\,n\,(d+1)^n$, or,
 for the case simulated, 48.  
Besides the exponential increase of dimension with `$n$', the only available simulation
routine (that of Tao; see Statistical and Computational Appendix 
for discussion and references) requires a great
number of time-steps, which may scale with the size of the system.\footnote{Tao was aware
of this limitation to his method, see \cite{Tao} and \cite{WickIII}. Nevertheless,
it's the only game in town if you want to simulate a nonseparable, nonlinear Hamiltonian system.} 

The runs necessitated 
20,000 time-steps for numerical stability (preserving energy and norm),
requiring about 3 1/2 minutes each on a PC. The study that produced Table 2 took around
10 hours. If you would prefer I had 
used $n=4$, the dimension 
would have been 128---but the algorithm may not scale linearly with dimension. 
Also, stability required 80,000 time-steps per run. The study
would have taken at least 120 hours, but testing indicated much more.
Since I needed to repeat the study
many times to experiment with parameter ranges, this was impractical. 

If the reader happens to have a supercomputer at hand, please do extend this study.  

\bQ What's the point of introducing the model, anyway? Why not just state your case and leave it at that.

\bA When responding to a claim that
something is not possible, ``But it is!" doesn't have much force.
``Look at this output!" is a better riposte.

\bQ The reliance on computer simulations worries me.

\bA Me, too; also, the results reported here came after many hours of code-writing, testing, and
experimenting with parameter ranges. I would prefer to have spent that time 
doing some traditional math.
 But no one can solve high-dimensional, non-linear
equations on their yellow pad! That reality is one of the reasons we now have access to
electronic computers. (Both John von Neumann in the U.S. and Alan Turing in the U.K.
were frustrated by their inability to solve such equations, von Neumann even remarking
that ``the analytic approach to nonlinear problems has failed." In one of his last
acts before biting into the cyanide apple, Turing was programming the first 
commercially-available computer to solve reaction-diffusion equations 
which he hoped could explain morphogenesis
in biology.) 

\bQ Prove to me that your program actually solves, approximately, your model.

\bA I cannot. There is no known method for proving that your program does what you hope it will;
indeed, I suspect that no such methodology could exist. Even if the logic of your
code can be checked, there is always the issue of physical implementation. 
Does your chip have hidden errors? \footnote{I recall an episode where
somebody discovered that their chip did not perform floating-point multiplication
correctly.} 

Besides such a sobering
lesson from the computer scientists,
I also cannot assure you that no bugs remain in the code. Thus output from a 
computer must be regarded as like outcomes from laboratory experiments: some evidence that your
theory is working, or that it is false.

\bQ Doesn't your model incorporate some kind of action-at-a-distance? To explain
why only one detector gets triggered (in some runs)? So your model can't be relativistically
invariant.

\bA True. It is based on \Sch's equation of 1926, which was not relativistically-invariant.
But there are actually two choices for WFE having the right properties: one restricting
spatial dispersion of the center-of-mass, 
and another restricting dispersion of the ``center of momentum".
In the first paper, \cite{WickI}, I
 argued that the latter version can make for a special-relativistically-invariant
theory, along the lines proposed by Dirac. In a later paper, \cite{Wickspacetime}, 
I argued that a nonlinear wavefunction theory can even be made compatible with
General Relativity (with a generalization of the notion of wavefunction, and 
some mild restrictions on allowable space-times).

\bQ I don't like your use of p-values. Weren't they invented for use in agriculture or medicine,
and lately suspect even in those fields? Besides, I don't want to know that I can't
reject the Born hypothesis (on the basis of some artificial data); 
I want to know what is the probability that it is true!

\bA Indeed, in the 1920s the English statistician and evolutionary theorist 
R. A. Fisher introduced p-values
as a tool for assessing the significance of a data set in supporting some hypothesis; 
it is defined as the probability that
the interesting observation you have made was just a fluke of chance.\footnote{More
precisely, the p-value is the probability that, under the ``null hypothesis", the 
quantity of interest can be as extreme as that observed.} Almost as an aside, Fisher remarked
that .05 would be a useful criteria for significance. 
But late in life he expressed regret for ever mentioning it. 
The rule adopted in the biomedical sciences: ``Get a .05 and you publish!" has
lead to the ``reproducibility crisis" haunting many fields.

I, too, wish I knew the probability that Born's hypothesis, or for that matter, my theory,
is true. But to my knowledge no statistician, probabilist, or philosopher of science has ever made
sense of such a statement. Doing so would require knowledge of the whole universe of possible
alternative theories, which no one can ever know.  

\bQ You say you ``experimented" (messed around?) with parameters until you found
your Dot on the Screen. Isn't that a case of falling into the Etch-a-Sketch Trap,
that you once decried in a book (\cite{Wickfittingbook}, section 7.1)?

\bA I referred in that book to mathematical modelers working in biology who
built models with hundreds of parameters, which they may have fixed by a computer
search for ``best fit to the (always very limited) data set". That is a sin. But in this work
I had to adjust only seven parameters: `$w$', `$v$',`$\alpha$', the two masses, `$\delta$',
and the final time.
If you possessed apparatus (say, a photographic plate  or ``photon detectors") two of
those parameters would already be known, as well as the masses (or perhaps be irrelevant;
also, only some combinations, e.g., $v$/DMass, may be relevant).
While in biology, usually most parameters are unknown and unmeasureable (e.g.,
in pandemic modeling, what's the probability that people will agree to mask up?)

I would also claim I am guilty only of the sort of selectivity routinely 
practised by experimentalists:
e.g., those who saw That Dot on the Screen were, perhaps by chance, those whose apparatus 
was permissive of that outcome, while others may have been actively searching for it by adjusting
fluxes, detector settings, etc.
 
\bQ In particular, the purpose of your parameter `$\delta$' seems to be to create 
a dichotomy to replace a continuum of ``dots". 

\bA It is a common occurence in many scientific and engineering 
contexts that an analogue signal gets
 converted into 
a binary one. Often a threshold exists defined by a minimum energy to be supplied, e.g.,
in photoreception in the eye, and likely in artificial ``photon detectors". Or it can be a minimum
electrical potential, as is necessary for a neuron to fire. But must it be true that
``dots on a screen" are an all-or-nothing phenomenon? Or could there be shades of darkening?  
Perhaps a certain minimum contrast in necessary before your eye can detect a dot.

\bQ In the statistical analysis, you culled the data to single out the single-dot runs, 
ignoring the rest. Isn't that cheating? 

\bA No. I merely imitated what an experimenter will certainly do who is interested solely in
\TDotS. Similarily, an epidemiologist analyzing transmission of infection in a population 
will ignore all the cases where there was no transmission, or no infection. Even Millikan
culled all the drops in his oil-drop experiment that ``misbehaved" (by failing to be charged or
for other reasons). 

\bQ You wrote, in section \ref{DaCsection},
``Let's set $d = 1$, so that the detectors can take only two values:
$y_j = 0$ means $j$-th detector untriggered, and $y_j=1$ means it was triggered."
So why isn't that what we see, given that the `detector' is a macroscopic device?

\bA In strict \Schism, nothing is real besides the wavefunction.\footnote{Excluding gravity.
I accept that Einstein got it right that gravity is space-time curvature and that
his metric tensor and curvature tensor represent elements of reality. Wavefunctions
describe matter, and should appear on the right-hand side of his gravity equation of 1915.} 
So the observable
can only be given by some functional
of the wavefunction. To declare that {\em both} $\psi$ {\em and} $\uy$ exist separately,
would be to agree with Bohr that there are separate realms, one ``quantum" and described
by a wavefunction, the other ``classical" and given an ordinary description.

\bQ You found that, if WFE is omitted, the observable and the Born probabilities were
``almost flat". So no dot. What's the story here?

\bA The original motivation for proposing WFE in wavefunction physics was to
recover the ``classical world" at large scales and normal energies---meaning
ordinary observations, e.g., the pointer on the dial moved to the left, or the 
Geiger counter clicked, or \TDotS\, appeared.  
In this toy model,
adding a large-enough WFE has a localizing effect which, in \Copism, would
be attributed to the particle hitting the screen at that spot.
So WFE is doing its job.

\bQ Even if your dot appears, won't it go away?

\bA Probably. The model is Hamiltonian and, assuming incorporation of the nonlinearity,
may well be ergodic. If so, the system is guaranteed to eventually recure to the
initial (dotless) state. I only claim that if you make a snapshot of the screen,
you may see a single dot. The question of how a system can stabilize a change-of-state,
like, e.g., a registration in a computer memory, involves the thermodynamic limit
and the issue of how dissipation can arise from a reversible system, 
matters beyond our purview here.
 
\bQ A wave-plus-particle (Bohmist), or a wave-until-you-measure-something (\Copist),
 interpretation plus Born's probability hypothesis
works well for all practical purposes.\footnote{Or: {\em FAPP}, as Bell abbreviated it, 
\cite{Belllastpaper}, and asked whether that is good enough 
for the supposedly fundamental theory of everything.} 
Why do we need a new, nonlinear, theory, so much more difficult to
solve?

\bA For several reasons concerned with ideology or philosophy, and because it is testable
and likely to have (unforeseeable) consequences. For the first, I regard it as part of a program
to drive mystical or metaphysical excuses out of physics, and replace them by better equations.
For the second, it serves as a reminder that, for any explanation of phenomenon, however
intuitive or established, there are always alternatives. 
Finally, if WFE
exists, then there are thresholds in physical systems, \cite{WickIII}, 
\cite{Wickchaos}, \cite{WickIV},
that as yet have not been explored.

\section{Discussion\label{discussionsection}}

I can imagine two scenarios in which That Dot on the Screen might appear. First, in one
of many repetitions of an experiment, which happens to hit a lucky initial condition.
Second, in a single experiment allowed to run; at some time, the dot briefly appears.
The latter scenario might rely on another proposed characteristic of Hamiltonian chaos:
``topological transitivity", meaning that the system will over time sample every
configuration consistent with energy conservation and other constraints. Thus, if the 
dot is allowable, it will someday appear. 

In paper \cite{Wickchaos}, I demonstrated \sdoic\ in a different model by simulation; of course,
doing so for topological transitivity is impossible. Note that the second
scenario would implicate a time-scale very difficult to predict.

Let's return to the topic of chance in physics. 
Recall \Poin's discussion of why, and in which situations, 
we find the probability concept useful. 
The canonical examples are a coin-flip and roulette. Both are (ignoring
air currents in the room) completely deterministic.
But both set-ups are designed so as to expand and smooth 
out any randomness in the initial conditions (the angular-velocity of the coin, plus the distance
to the table or wrist; the croupier's
toss, plus the orientation and speed of the wheel). Thus for each we can adopt the law:
all outcomes equally likely. Now glance again at the histograms in Figure 3 of dot locations
(on runs where only one detector triggered).
It appears that the distribution of dot positions is getting spread out, but bypassing
location 2.0; perhaps the system is imitating the coin rather than the roulette wheel.
Or, if I could greatly increase `$n$', or run the process
much longer, the histograms would flatten out and I would have created 
a sort of crazy roulette game.

How did Born fare in this study? As remarked in section \ref{simsection}, there
were arguments for-and-against rejection of Born's hypothesis. However, the
model is so limited in scale and realism that I will proclaim myself agnostic on this issue.
 
Finally, I would like to point out that, despite the language sometimes appearing 
between quotation marks,
there are no ``particles" in this model. The word is still useful; but in this context
it refers to \TDotS\ and nothing more. In other words, in wavefunction physics,
particles are mere epiphenomena (phenomena appearing in conjunction with other such,
here meaning the detectors, whatever they may be).

\section{Appendix: Statistical and Computational Issues\label{statandcompsection}}

Imagine that you have a set of 100 dice.\footnote{``dice, n.pl. [sing. die, dice] 
1. small cubes of bone, plastic, etc. marked on each side with a different number of spots 
(from one to six) and used
in games of chance." Webster's New World Dictionary of the English Language, College Ed.,
{\em circa} 1968.} The dice are variously weighted, and,
for each die you possess a list of the probabilities for showing each number on a roll,
say $\{p_{i,r}\}$, where $i = 1,2,..,6$ and $r = 1,2,...,100$. Now someone presents with
a sequence of single-die outcomes, say $\{\mu(r)\}$, where $\mu(r) \in \{1,2,...,6\}$ for
$r = 1,2,...,100$. Morever, this person claims these outcomes resulted from
rolling your dice set, once with each die. Question: how plausible is this claim?

One approach, which I used for the simulations, is Fisher confidence testing, adopting
the weight list as the so-called ``null hypothesis". Fisher's method is to construct
a statistic which I will call `$S$' using the data, and then calculate the probability
that this statistic would come out as ``extreme" as the observed $S$ if the claim is true,
i.e., the data was derived as claimed from your dice set. Reject the claim if this
so-called ``p-value" is small, e.g., $p < 0.05$. (For biologists. For physicists,
if $p < 0.00001$, i.e., ``five sigma".)

The statistic should be a measure of ``extreme-ness" in some sense. A good choice\footnote{The
negative log likelihood; probably the optimal choice.} is

\be
S \= - \sum_{r=1}^{100} \, \log(\,p_{\mu(r),r}\,).
\ee 

In the application, `$r$' represents a run of the model in which exactly one detector
was judged to be triggered (i.e., $\mu(r) = i$ if $\Ob_i > \delta$ for this `$i$' alone),
and $p_{i,r} = \Born_{i,r}$. In other words, the statistical test reflected how likely it
is that the observed triggered sites were derived from the Born probability vector 
for the same run as if a die with `$n$' sides and those weights were thrown. 

These probabilities can be approximated using, e.g., central-limit theorems and so forth.
But nowadays, with speedy electronic computers available, it is easier and more accurate
just to simulate the experiment say, 100,000 times and count. I checked
that the routine gave roughly the correct ``rejection fraction".\footnote{Note 
that I have not defined an ``alternative hypothesis".
This is a pecularity of Fisher's method that renders it useless for theory choice:
the alternative hypothesis is irrelevant for the test, and the assertion that, if you can reject
the null hypothesis, then the alternative must be true is a {\em nonsequitur} (except in the 
implausible situation that only two theories are conceivable).}

The statistic `$S$' is large if the observed numbers came from dice for which the weights
assigned the observed sides low probability. Thus the test rejects if too-many low probability
events occurred in the sample. But what if the sample outcomes were ``too probable" in a sense?
For example, suppose for each die the weights had a unique maximum, and the sample
just happened to hit each such side; obviously we should then reject the null. An
appropriate statistic then would be:

\be
R \=  \sum_{r=1}^{100} \, \left[\,\mu(r) - \hbox{maxp(r)}\,\right]^2,
\ee 
  
\ni where $\hbox{maxp}(r)$ is the maximum Born probability on that run.

The model was simulated using the Tao method, \cite{Tao}, \cite{WickIII}: 
a direct, symplectic solver that works
for nonseparable Hamiltonians (the only such known to the author). Programs were
written in the C language and run on a 15-year-old HP PC. See also \cite{Wickchaos}
for how determinants were computed.


\begin{thebibliography}{9}

\bibitem{Mott}
Mott, N. F. ``The Wave Mechanics of $\alpha$-Ray Tracks". {\em proc. Royal Soc. London}.
A126, 79-84 (1929), 
reprinted in 
{\em Quantum Theory and Measurement}, Wheeler, J. A. and Zurek, W.H., eds., 
Princeton University Press (1983). The ``Red Book".


\bibitem{Born}
Born, M. ``Zur Quantenmechanik der Stossvorg{\" a}nge". {\em Zeitschrift f{\" u}r
Physik}, 37,863--67 (1926), 
reprinted and translated into English in the Red Book.

\bibitem{EPR}
Einstein, A., Podolsky, B., and Rosen, N. ``Can Quantum-Mechanical Description of
Physical Reality Be Considered Complete?" {\em Physical Review}, 47, 777-80 (1935),
Reprinted in the Red Book.

\bibitem{catpaper}
\Sch\ E. ``Die gegenw{\" a}rtige in der Quantenmechanik". {\em Naturwissenschaften} 23: 807 (1935);
reprinted and translated into English in the Red Book.

\bibitem{vonNeumann}
Von Neumann, J. {\em The Mathematical Foundations of Quantum Mechanics}. 
English translation of the original German text (1932), 
Princeton University Press, 1955. 

\bibitem{Bell}
Bell, J. S. {\em Speakable and unspeakable in quantum mechanics}. Cambridge University Press 
(1987; paperback edition 1988). See article 18.

\bibitem{WickI}
Wick, W. D. ``On Non-Linear Quantum Mechanics and the Measurement Problem I: Blocking Cats", 
ArXiv 1710.03278 (October 2017).

\bibitem{WickII}
Wick, W. D. ``On Non-Linear Quantum Mechanics and the Measurement Problem  
II: The Random Part of the Wavefunction", ArXiv 1710.03800, (October 2017).  

\bibitem{WickIII}
Wick, W. D. ``On Non-Linear Quantum Mechanics and the Measurement Problem  
III: Poincar{\' e} Probability and ... Chaos?" ArXiv 1803.1126v1 (March 2018).

\bibitem{WickIV}
Wick, W.D. ``On Non-Linear Quantum Mechanics and the Measurement Problem IV: Experimental Tests".
ArXiv 1980.02352v1 (6 August 2019)


\bibitem{Wickchaos}
Wick, W. D. ``Chaos in a Nonlinear Wavefunction Model: An Alternative to Born's Probability
Hypothesis". ArXiv 2502.02698v. (4 Feb. 2025).

\bibitem{Poincare}
Poincar{\" e}, H. {\em Calcul des Probabilit{\" e}}. 1896.

\bibitem{Tao}
Tao, M. ``Explicit symplectic approximation of nonseparable Hamiltonians: 
algorithm and long-time performance." Arxiv 1609.02212v1, September 2016; 
Phys Rev E. 94: 043303 (2016).


\bibitem{Wickspacetime}
Wick, W.D. ``On Non-Linear Quantum Mechanics, Space-Time Wavefunctions, and Compatibility
with General Relativity". ArXiv 2008.08663v1. (19 August 2020).


\bibitem{Wickfittingbook}
Wick, W.D. {\em Fitting Non-linear, Stochastic Models to Data in Biology and Medicine}.
Available from Amazon, print or e-book (2012).

\bibitem{Belllastpaper}
Bell, J. S.  ``Against `measurement' ''. Proceedings of 62 years of Uncertainty, 
a conference held in Erice, Italy, 5-14 August 1989, 
published by Plenum Publishing, New York; reprinted in {\em Physics World}, August 1990, p. 33-40. 




\end{thebibliography}
\end{document}